\documentclass[aps,prl,superscriptaddress,floatfix]{revtex4}

\usepackage{graphicx}
\usepackage{epsfig}
\usepackage{amssymb}
\usepackage{amsmath}

\newcommand{\dd}{\mathrm{d}}

\begin{document}

\title{Nonlinear NMR dynamics in hyperpolarized liquid $^3$He}

\author{E.~Baudin}
\affiliation{Laboratoire Kastler Brossel, Ecole Normale
Sup{\'e}rieure; CNRS; UPMC; 24 rue Lhomond, F75005 Paris, France}
\author{M.~E.~Hayden}
\affiliation{Physics Department, Simon Fraser University, 8888
University Drive, Burnaby BC, Canada V5A 1S6}
\author{G. Tastevin}
\author{P.~J.~Nacher}
\affiliation{Laboratoire Kastler Brossel, Ecole Normale
Sup{\'e}rieure; CNRS; UPMC; 24 rue Lhomond, F75005 Paris, France}

\begin{abstract}
\noindent{\bf Abstract}
\vskip 0.5\baselineskip
\noindent
In a highly polarized liquid (laser-polarized $^3$He-$^4$He mixtures in our experiment), dipolar magnetic interactions within the liquid introduce a significant nonlinear and nonlocal contribution to the Bloch equation that leads to instabilities during NMR evolution. We have launched a study of these instabilities using spin echo techniques. At high magnetizations, a simple 180$^{\circ}$ rf pulse fails to refocus magnetization, so we use a standard solid-state NMR pulse sequence: the magic sandwich. We report an experimental and numerical investigation of the effect of this sequence on unstable NMR evolution. Using a series of repeated magic sandwich sequences, the transverse magnetization lifetime can be increased by up to three orders of magnitude.

\vskip 0.5\baselineskip

\noindent{\bf R\'esum\'e}
\vskip 0.5\baselineskip
\noindent
Dans un liquide fortement polaris\'e (ici, une solution d'$^3$He hyperpolaris\'e dans l'$^4$He liquide), les interactions dipolaires magn\'etiques apportent une importante contribution non lin\'eaire et non locale \`a l'\'equation de Bloch qui rend l'\'evolution RMN instable. Nous avons entrepris une \'etude de cette instabilit\'e par des techniques d'\'echo de spin. A forte aimantation, une impulsion rf de 180$^{\circ}$ appliqu\'ee lors de l'\'evolution ne refocalise pas l'aimantation, aussi utilisons nous une technique standard de RMN du solide : le \lq \lq magic sandwich\rq \rq . Nous d\'ecrivons les premiers r\'esultats d'une \'etude exp\'erimentale et num\'erique des effets de cette s\'equence sur une \'evolution RMN instable. Une augmentation de la dur\'ee de vie de l'aimantation transverse de 3 ordres de grandeur est obtenue en appliquant p\'eriodiquement des \lq \lq magic sandwich\rq \rq .

\vskip 0.5\baselineskip
\noindent Accepted for publication in C. R. Chim. (2007), doi:10.1016/j.crci.2007.07.005
\end{abstract}

\maketitle

\section{Introduction}
Conventional treatments of nuclear magnetic resonance (NMR) ignore collective effects for spin dynamics, and thus the evolution of the local magnetization of a single spin species is described by simple (linear) Bloch equations. This approach is usually justified in liquids, where short-range dipolar spin interactions (averaged out by fast atomic diffusion) only lead to relaxation. But it is actually limited to conventional implementations of NMR where equilibrium magnetizations are very small. The emergence of high-resolution NMR spectroscopy as a powerful tool for condensed matter physics, analytical chemistry, structural biology, and medicine, has motivated considerable efforts to improve signal-to-noise ratios through increased use of high static fields and of probes with high quality factors. These approaches enhance nonlinear contributions to spin dynamics that 
arise from radiation damping \cite{vlassenbroek95}. A more complex situation arises when additional nonlinear contributions of long-range dipolar interactions between the local magnetization and that of the remainder of the sample are considered. The influence of these dipolar interactions becomes significant at large magnetization densities and is thus expected to play an increasing role as technological advances lead to the use of higher and more uniform static fields. In particular, they are expected to lead to spin turbulence and spatiotemporal chaotic behavior \cite{lin00,datta06}.

Spectacular dynamical effects resulting from distant dipolar fields (DDF) have been observed in liquids obtained by condensation of hyperpolarized $^3$He (i.e. having a high, out-of-equilibrium nuclear polarization produced by laser optical pumping). These effects range from spectral clustering (at small tipping angles, with the spontaneous appearance of long-lived geometry-dependent modes of coherent nuclear precession \cite{candela94,jeener99,sauer01,ledbetter02}), to precession instabilities (at large tipping angles, with exponential growth of spatially inhomogeneous magnetization patterns \cite{nacher02,huang02}). These effects have been accounted for, at least qualitatively, by theoretical approaches and numerical studies \cite{jeener02}. 
Experimental investigations of nonlinear DDF effects are challenging with thermally polarized liquids; the characteristic dipolar field for $^1$H at 12~T is 43~nT. In contrast, dense laser-polarized systems such as liquefied $^3$He are ideal model systems for the study of nonlinear NMR dynamics. $^3$He has a high nuclear magnetic moment, and its liquid phase provides dipolar fields as large as 100~$\mu$T for pure $^3$He polarized to 45\%. 

We have recently succeeded in generating spin echoes in highly polarized $^3$He-$^4$He dilute mixtures (with dipolar fields of order 1~$\mu$T), and have demonstrated that it is possible to dynamically stabilize magnetization distributions in intrinsically unstable regimes \cite{hayden07}. 
Here, we report additional experimental results, provide a discussion of echo decays both at low and high DDF, and present a comparison to results of exploratory numerical simulations.

\section{Experimental setup}

\subsection{Preparation of hyperpolarized helium samples}
The $^3$He nuclei in our experiments are hyperpolarized by optical pumping, which amounts to the transfer of angular momentum from an intense circularly polarized laser beam to atoms by light absorption.
$^3$He gas is continuously injected at low pressure (mbar) into the room temperature part of a glass cell where it is optically pumped in a few seconds to 40\% polarization by metastability exchange optical pumping \cite{colegrove63}, using a 2~W fibre laser at 1083~nm \cite{tastevin04}.
The polarized gas flows down a narrow tube into the low temperature (1~K) part of the
apparatus (a 0.44~cm$^3$ spheroidal Pyrex cell) partially filled with superfluid liquid $^4$He.
Wall relaxation is inhibited through the use of Cs coating on the glass, and bulk dipole relaxation is of order several hours at the low $^3$He molar fractions used in our experiments (typically 1-5\%) \cite{piegay02}. 
We control the volume of liquid admitted to the cell (typically half full) and the nuclear polarization in the liquid phase (up to 40\%). We can thus choose the initial magnetization $M$, and the $^3$He spin diffusion coefficient $D$ (in the $10^{-3}-10^{-2}$~cm$^2$/s range, controlled both by the temperature $T$ and the $^3$He molar fraction $X$) for each experiment. 
At large tipping angles, NMR excitation induces a significant irreversible loss of this out-of-equilibrium magnetization, so that a new batch of polarized gas has to be condensed for each experiment.  

\subsection{NMR setup}
\label{setup}
NMR is performed at very low magnetic field (2.3~mT, shimmed to 20~ppm over the sample) using crossed coils. 
The frequency of the rf pulses (74.5~kHz) is set to lie within 1~Hz of the Larmor frequency corresponding to the average of the external field $\vec B_{\textrm{ext}}$ over the sample.
The magnetic field is produced over the entire apparatus by a series of seven coils. 
Slow fluctuations of $\vec B_{\textrm{ext}}$ due to remote magnetic perturbations are measured with a fluxgate magnetometer, and compensated using additional bucking coils. Overall field fluctuations are thus reduced by more than a factor of 10, and the corresponding changes in Larmor frequency  are found to be smaller than 0.5~Hz over the duration of experiments (several minutes). 

The rf transmit coil is actively shielded \cite{bidinosti05} to prevent eddy currents from significantly heating the low temperature parts of the apparatus when an oscillating rf field $\vec B_{\textrm{rf}}$~($\simeq 30$~$\mu$T) is applied. 
This also ensures that eddy currents do not distort the applied rf field, that is uniform to within $±200$~ppm over the volume of the sample cell.
The receive coil is tuned to the Larmor frequency with a very low Q factor ($\simeq 2$) so that radiation damping effects are negligible. 
In-house PC-controlled analog hardware allows flexible control of hard pulse sequences.

\section{NMR evolution in a magnetized liquid}
We use the following modified Bloch equation to describe the NMR evolution of local magnetization $\vec M$ in the frame rotating at the rf angular frequency: 
\begin{equation}
\frac{\dd \vec M}{\dd t} = \gamma (\vec B_{\textrm{eff}}+ \vec B_{\textrm{DDF}}) \times \vec M  + D \Delta \vec M,
\label{Blocheq}
\end{equation}
where $\gamma$ is the $^3$He gyromagnetic ratio ($\gamma/2\pi=32.4$~MHz/T) and $\vec B_{\textrm{eff}}$ is the effective field resulting from $\vec B_{\textrm{ext}}$ and $\vec B_{\textrm{rf}}$. $\vec B_{\textrm{DDF}}$ is the distant dipolar field induced by the distribution of magnetization. 
The conditions of our experiments are such that contributions to the Bloch equation from relaxation and radiation damping can be neglected. 

Using a secular approximation in the rotating frame, the distant dipolar field can be expressed as:
\begin{equation}
\vec B_{\textrm{DDF}}(\vec r) \!=\!\frac{\!\mu _{0}}{8\pi }\int \!\! \dd \vec r' \frac{2 \vec M_{z}(\vec r')-  \vec M_{\perp}(\vec r')}{\mid \vec r- \vec r' \mid ^3}(3\cos^2\theta-1),
\label{DDField}
\end{equation}
where $\theta$ is the angle between $\vec r - \vec r'$ and the average direction $\hat z$ of $\vec B_{\textrm{ext}}$.
$\vec B_{\textrm{DDF}}$ brings a nonlinear nonlocal contribution to the Bloch equation (Eq. \ref{Blocheq}). 

The main features of the dynamical evolution of magnetization can be discussed in terms of three parameters: $\gamma \Delta B_{\textrm{ext}}$, the rate at which the external field inhomogeneities defocus magnetization ($\Delta B_{\textrm{ext}}$ being a measure of the typical variation of $\vec B_{\textrm{ext}}$ over the sample); $\gamma B_{\textrm{dip}}$, a characteristic rate associated with the initial DDF ($B_{\textrm{dip}}$=$\mu_{0} M$ is the amplitude of the dipolar magnetic field created by homogeneous magnetization $\vec M$ before the first rf pulse of the sequence); $D/r^2$, the diffusion-induced damping rate for a magnetization pattern of typical scale $r$. 
It is important to note that $\vec B_{\textrm{DDF}}$ is the field created by the instantaneous distribution of local magnetization and evolves in both space and time, whereas $B_{\textrm{dip}}$ is fixed before each NMR experiment. 

\section{DDF-induced NMR precession instabilities}

\begin{figure}[ht]
\begin{center}
\includegraphics[width=10cm]{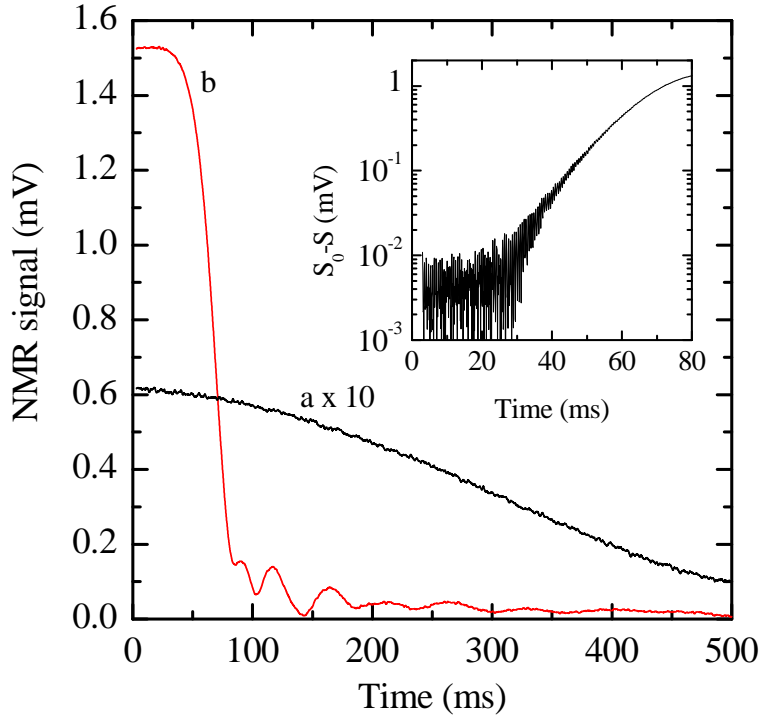}
\caption{\label{Exponential} FID response of laser-polarized $^3$He in liquid $^4$He to a 45$^{\circ}$ tipping pulse for $B_{\textrm{dip}}$=51~nT (negligible dipolar interactions, trace a) and a 90$^{\circ}$ tipping pulse for $B_{\textrm{dip}}$=0.9~$\mu$T (trace b). The inset shows the deviation of trace b from its initial value $S_{0}$, in semi-log scale. Experiments are performed on the same sample at $T=1.14$~K, with $X\simeq2\%$}  
\end{center}
\end{figure}

Figure \ref{Exponential} displays examples of free induction decay (FID) signals following large tipping angle pulses in liquid $^3$He-$^4$He mixtures. 
At low magnetization density (negligible dipolar interactions), we observe a characteristic FID half-life of 0.3~s, a value that is set by the homogeneity of the magnetic field over the cell (Fig. \ref{Exponential}, trace a). 
In contrast, an experiment performed with $B_{\textrm{dip}}=0.9\mu$T reveals a dramatic collapse of the FID signal on a timescale of 70 ms (Fig. \ref{Exponential}, trace b).
The deviation of the NMR signal $S$ from its initial value $S_{0}$ grows exponentially  with time (inset to Fig. \ref{Exponential}).

A model developed by J. Jeener \cite{jeener99} and numerical simulations described in section \ref{Dipolar} \cite{datta06,nacher02,theseMarion} have shed light on magnetization dynamics following a large tipping angle pulse: 
at large $B_{\textrm{dip}}$, during transverse precession, unstable inhomogeneous magnetization patterns develop and grow exponentially with time. 
These patterns can arise from an initial seed of inhomogeneity following the tipping pulse, from stray or applied static field gradients, or from edge effects that depend on the sample shape.
Jeener's model for infinite media predicts that the instability growth rate $\Gamma_{\textrm{S}}$ of the deviation $S_{0}-S$ is proportional to $B_{\textrm{dip}}$. 
This analytical model and corresponding numerical simulations yield the value $2\pi\Gamma_{\textrm{S}}/\gamma B_{\textrm{dip}} = 2 \sqrt{2} \pi/3 \simeq 3.0$. 
Simulations of finite-sized samples and previous experiments on spheroidal samples agree with one another, but yield a significantly lower ratio $2\pi\Gamma_{\textrm{S}}/\gamma B_{\textrm{dip}} \simeq 2.3$ \cite{nacher02,theseMarion}. The present experiments confirm the latter result. 

\section{Time-reversal experiments using magic sandwich sequences}

Our objective is to characterize the onset and growth of DDF-induced inhomogeneous magnetization patterns, beyond the mere measurement of a global growth rate. In particular, we are interested in the determination of spatial length scales associated with magnetization patterns that develop during unstable NMR precession. 
To this aim, we plan to use echo techniques and the established link between diffusion-induced NMR relaxation rates and dominant frequencies of spatial modulations of the magnetization density. 
The conventional Hahn echo sequence ($90^{\circ}-\tau-180^{\circ}$) effectively generates a spin echo in weakly magnetized systems, such as laser polarized gas or dilute liquid mixtures as long as the nuclear polarization is kept very low. In contrast, this sequence fails completely when used for experiments on highly magnetized mixtures \cite{nacher02,hayden07}. 
This failure is due to the nature of the dipolar interactions; the 180$^{\circ}$ pulse reverses both $\vec M$ and $\vec B_{\textrm{DDF}}$, leaving the quadratic term in Eq.~(\ref{Blocheq}) unchanged. 

The Magic Sandwich sequence (MS) was introduced during the 1970s in the context of
solid-state NMR to compensate for the strong effect of local dipolar fields \cite{rhim71}. 
In our experiment it has been adapted and used, for the first time, on a liquid system. 
It consists of continuously applied and appropriately phased rf pulses during a given
time interval. 
It mixes longitudinal and transverse components of $\vec M$, that contribute in a different way to the dipolar field (Eq.~(\ref{DDField})), and causes a time-reversed evolution for the magnetization at half the pace of the free evolution while it is applied. Applying quasi-cw rf for up to several seconds without significantly heating
the system is made possible through the use of shielded coils and by operating at very low NMR frequency. 

\subsection{Effect of rf detuning}
\begin{figure}[ht]
\begin{center}
\includegraphics[width=8cm]{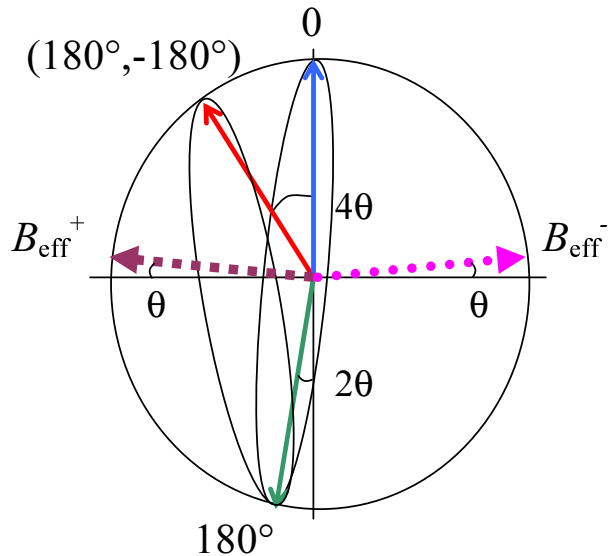}
\caption{\label{Bloch} Angular error associated with rf detuning. 
In the rotating frame, the effective rotation axes are tilted from the $\hat x$-axis by an angle $\theta$, proportional to the rf frequency offset. 
After the first 180$^{\circ}$ pulse, the Bloch vector lies $2\theta$ away from the targeted $\hat z$-axis, and the $(180^{\circ}_x,-180^{\circ}_x)$ pulse pair leads to a net $4\theta$ angular tilt. Here, $\vec B_{\textrm{DDF}}$ is assumed to have a negligible effect. 
}
\end{center}
\end{figure}

The magic sandwich sequence introduced by Rhim, Pines and Waugh \cite{rhim71} can be summarized as: 
\begin{equation} 
90^{\circ}_y,(180^{\circ}_x,-180^{\circ}_x)_n,-90^{\circ}_y, 
\label{MSeq}
\end{equation} 
where the 90$^{\circ}$ and 180$^{\circ}$ rotations are performed along orthogonal directions $\hat y$ and $\hat x$, and the $(180^{\circ}_x,-180^{\circ}_x)$ pulse pair is repeated $n$ times. 
This scheme only works for perfectly resonant applied rf field. 
A detuning $\Delta \omega$ between the Larmor and rf angular frequencies introduces a small angular tilt $\theta= \Delta \omega /\gamma B_{\textrm{rf}}$ between the desired rotation axis ($\vec B_{\textrm{rf}}$) and the effective rotation axis ($\vec B_{\textrm{eff}}$). 
The net result is that a (180$^{\circ}$,-180$^{\circ}$) pair will rotate a magnetization vector that is initially aligned with the $\hat z$-axis by an angle $4\theta$ (Fig. \ref{Bloch}). 

In our experiments, a typical MS sequence contains about 100 such rotation pairs and lasts 10-100~ms  (rf intensity is such that a 90$^{\circ}$ pulse lasts 0.5~ms). 
In order to avoid significant cumulated error ($\theta \ll \pi/1000$), one would need a small rf detuning ($\Delta \omega \ll 10$~rad/s). 
This cannot be achieved in our experiment due to the limited stability of the applied field $B_{\textrm{ext}}$. 
The sequence becomes robust against rf detuning when integer number of turns are performed, e.g. using $(360^{\circ}_x,-360^{\circ}_x)$ pulse pairs, and the direct (1$^{\textrm{st}}$ order) angular errors described above are eliminated. The remaining cumulative 4$^{\textrm{th}}$ order angular error has a negligible effect for $\Delta \omega = 10$~rad/s. This modification of the basic MS sequence (\ref{MSeq}) is necessary to ensure that it correctly reverses the effect of dipolar fields while the rf is applied.

\subsection{Experimental results}

\begin{figure}[ht]
\begin{center}
\includegraphics[width=11cm]{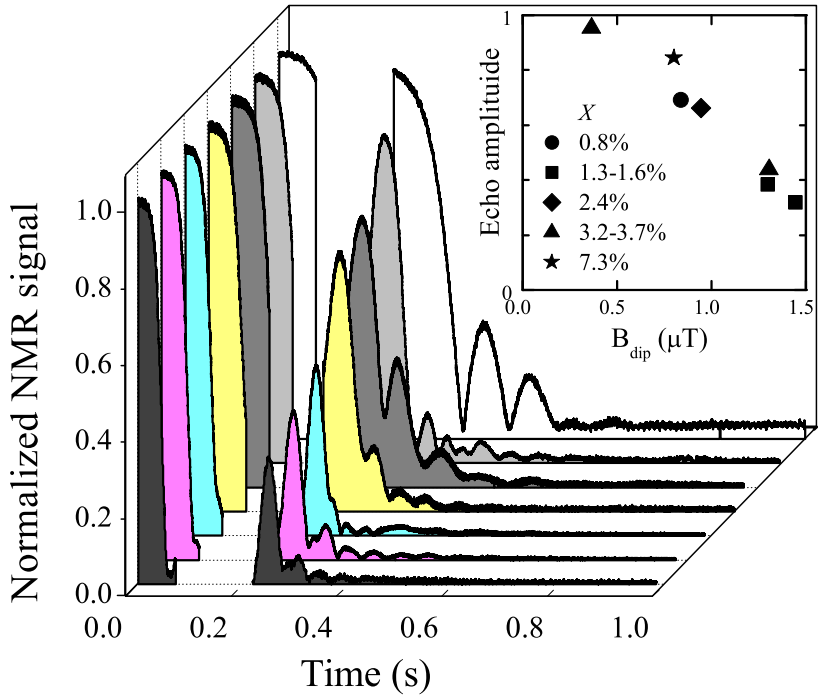}
\caption{\label{MS} Spin echo formation using a $90^{\circ}-\tau-\textrm{MS}$ sequence with MS duration $2\tau$=140~ms. Data are shown for increasing values of $B_{\textrm{dip}}$, from 0.3~$\mu$T to 1.5~$\mu$T (from back to front). FID signals have been normalized to facilitate comparison of data. Increasing $B_{\textrm{dip}}$ decreases the echo amplitude (inset), and significantly alters its shape. Experiments are performed at $T\simeq$1.14~K for the $^3$He molar fractions $X$ indicated in the legend. 
}
\end{center}
\end{figure}

Figure \ref{MS} displays the data obtained using a $90^{\circ}-\tau-\textrm{MS}$ sequence with MS duration $2\tau$,  performed for various magnetization densities. 
Magnetization freely evolves for a time $\tau$ during which the FID signal is recorded. Then a MS sequence is applied to unwind the magnetization back to time $t$=0, so that the subsequent free evolution is expected to yield an NMR echo signal identical to the initial FID signal. 

This type of behavior is observed at the lowest magnetization shown in Fig. \ref{MS}. At the highest magnetizations, echoes are still obtained long after the initial FID signals have vanished. 
However, the echo shapes progressively change, and their relative amplitudes decrease almost linearly with increasing $B_{\textrm{dip}}$ (inset to Fig. \ref{MS}). 
This decrease hardly depends on the diffusion coefficient, which roughly scales with the inverse of the $^3$He molar fraction over the investigated range. 
Imperfect refocussing can be attributed to sequence imperfections (in rf timing, phase or amplitude). 
Moreover, during the application of the MS sequence, while magnetization patterns that have developed during the free evolution period are driven back to their initial state, complementary families of patterns are expected to become unstable and grow \cite{jeener99}. 

The MS duration in the $90^{\circ}-\tau-\textrm{MS}$ sequence can be set arbitrarily to achieve time reversal and bring the dipolar-coupled system back to various times. 
In particular, when the MS duration is set to $4\tau$ (magnetization unwound back to $t=-\tau$) one observes an echo signal that peaks at time $t=6\tau$ (i.e., a maximal refocussing of magnetization at time $\tau$ after the rf pulses are applied). As recently reported, echo trains can be obtained using a series of such MS sequences, periodically repeated \cite{hayden07}. 
In the following section, we focus on a dedicated echo sequence, analyze the conditions for optimal efficiency in our experiment, and discuss echo attenuation rates. 

\section{Repeated magic sandwiches}

We use a Repeated Magic Sandwich (RMS) sequence: 
$$90^{\circ}-\tau-\textrm{MS}-2\tau-\textrm{MS}-2\tau-...$$
with MS cycles of duration $4\tau$. 
This is analogous to a conventional spin echo sequence in which the 180$^{\circ}$ pulses have been replaced by MS sequences. 
We show below that the RMS sequence can be designed to refocus the phase dispersions caused by both linear and dipolar couplings, and compare spin echo decay rates for RMS sequences and Carr-Purcell-Meiboom-Gill (CPMG) sequences \cite{meiboom58} in the limit of negligible dipolar interactions (Sect. \ref{attenuation}). RMS echo trains at high magnetizations are then described, and experimental and numerical results are compared (Sect. \ref{Dipolar}). 

\subsection{Repeated magic sandwiches with negligible dipolar interactions}
\label{attenuation}

\subsubsection{Expectations}

The magic sandwiches used for the RMS sequence are of the form: 
$$90^{\circ}_y, (360^{\circ}_x,-360^{\circ}_x)_n, 90^{\circ}_y.$$ 
Note that the sign of the last rotation has been changed with respect to the basic MS sequence (\ref{MSeq}) in order to remove the linear phase dispersion introduced by static field inhomogeneities (residual inhomogeneities or applied gradients). 
To experimentally study echo attenuation on time scales comparable to those relevant in dipolar-coupled systems, a static field gradient is applied to the weakly polarized liquid mixtures to induce controlled spatial modulation of the magnetization. 
During each MS, the intense rf field has a dominant contribution in Eq.~\ref{Blocheq}, and the evolution due to the gradient is frozen. The RMS sequence may thus be considered as being equivalent to a CPMG sequence in which the gradient is only applied between the rf pulses, i.e. for one third of the actual period of the RMS sequence (the period of the RMS sequence is $T_{\textrm{RMS}}=6\tau$, while that of the conventional CPMG sequence is $T_{\textrm{CPMG}}=2\tau$). 

Since spatial variations of $\vec M$ are damped by diffusion, echo amplitudes for both sequences are expected to decay exponentially in time with a rate: 
\begin{equation}
\frac 1 {T_2} = \frac{D k^2( \gamma G T_{\textrm{seq}})^2 }{12},
\label{Tanner}
\end{equation}
where $G$ is the amplitude of the gradient of the $z$ component of $\vec B_{\textrm{ext}}$, $T_{\textrm{seq}}$ is the period of the sequence, and $k$ a dimensionless factor \cite{stejskal65}. 
This factor is given by: 
\begin{equation}
k^2=\delta^2 (3 T_{\textrm{seq}}-2\delta)/{T_{\textrm{seq}}}^3,
\label{k}
\end{equation}
where $\delta$ is the effective duration of the applied gradient $G$. 
Hence, $k^2=1$ for the CPMG sequence ($\delta=T_{\textrm{CPMG}}$) and $k^2=7/27$ for the RMS sequence ($\delta=T_{\textrm{RMS}}/3$). A linear dependence of the decay rate on $(G T_{\textrm{seq}})^2$ is thus expected for both pulse sequences. 

\subsubsection{Experiments at very low magnetizations} 
A weakly polarized sample is easily prepared by addition of a small quantity of laser-polarized $^3$He gas to a pre-liquified unpolarized mixture. The added  $^3$He quantity ($\sim 0.5$~\% of the amount of $^3$He in the sample) is sufficiently small to have negligible influence on the $^3$He concentration, hence on the diffusion coefficient. 
Series of experiments can thus be performed on nearly identical samples.  

\begin{figure}[ht]
\begin{center}
\includegraphics[width=8.7cm]{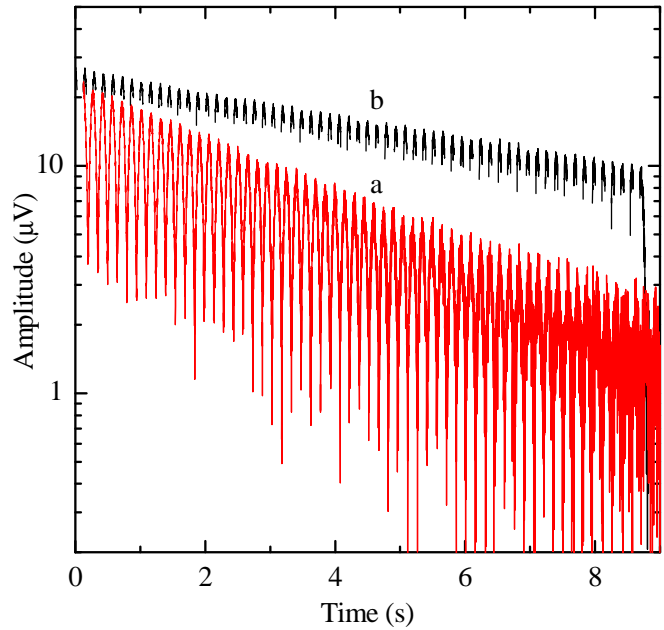}
\caption{\label{CPMG-RMS} 
Echo trains obtained under conditions where dipolar interactions are negligible using a CPMG sequence ($T_{\textrm{CPMG}}=151$~ms, trace a) and a RMS sequence ($T_{\textrm{RMS}}=145$~ms, trace b). 
Both experiments are performed on the same sample ($X$=1.9\%) at $T$=1.15~K with $G$=1.1~$\mu$T/cm. 
Decay rates $1/T_2$ of echo amplitudes are 0.28~s$^{-1}$ (trace a) and 0.12~s$^{-1}$(trace b). 
}
\end{center}
\end{figure}

Figure \ref{CPMG-RMS} shows echo trains obtained with RMS and CPMG sequences in the presence of an applied gradient. 
For both sequences, echo amplitudes are observed to decay exponentially with time. 
The decay rates $1/T_2$ are deduced from exponential fits on the squared echo amplitudes. This self-weighting of data reduces a systematic bias in the analysis associated with Rician noise in amplitude data \cite{guillot02}. 
In our preliminary investigations, systematic measurements with both sequences have not been performed on identical samples. 
\begin{figure}[hb]
\begin{center}
\includegraphics[width=8.7cm]{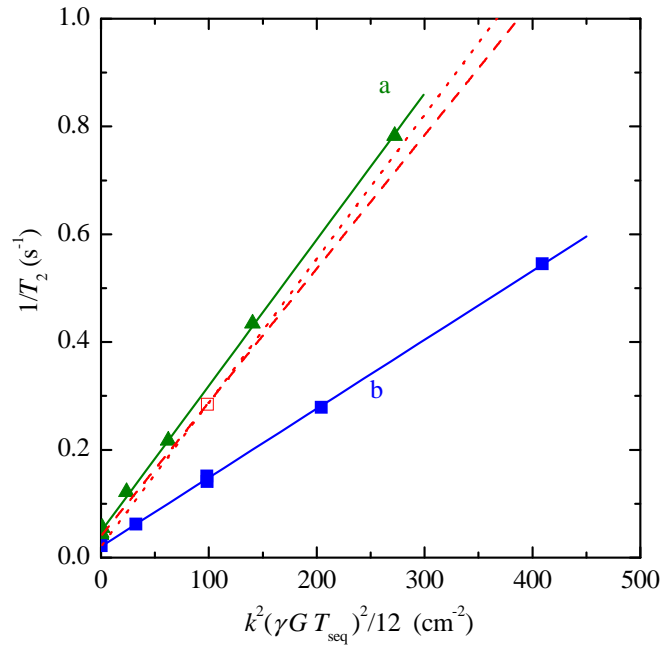}
\caption{\label{pentes} Comparison of RMS and CPMG decay rates under conditions where dipolar interactions are negligible. 
For sample~1 ($X$=1.9\%, $T$=1.15~K), RMS measurements are performed with $T_{\textrm{RMS}}$=145~ms and various gradient amplitudes (solid triangles), and a single CPMG measurement is performed with $T_{\textrm{CPMG}}$=151~ms at $G$=1.1~$\mu$T/cm (open square). 
For sample~2 ($X$=4.5\%, $T$=1.13~K), CPMG measurements are also performed with $T_{\textrm{CPMG}}$=151~ms (solid squares). 
These data are used to infer limiting values for CPMG decay rates in sample~1 (dashed and dotted lines, see text).
Solid lines a and b are linear fits to the measured RMS and CPMG decay rates, respectively. 
}
\end{center}
\end{figure}

Figure \ref{pentes} shows a compilation of RMS and CPMG decay rates measured in liquid mixtures with $^3$He molar fraction $X=1.9$\% (sample~1) and $X=4.5$\% (sample~2). 
Data are plotted as a function of the reduced parameter $k^2(\gamma G T_{\textrm{seq}})^2 /12$. 
Both sets of data exhibit the expected linear behavior in $(G T_{\textrm{seq}})^2$. 
The CPMG data (solid squares) reveal a small, but finite decay rate at zero applied gradient (imperfect shimming). 
The RMS decay rates (solid triangles) exhibit a larger intercept at $G$=0, and a more rapid increase with $(G T_{\textrm{RMS}})^2$. 
Much higher decay rates are indeed expected because of the lower $^3$He concentration, as confirmed by the CPMG measurement performed on sample~1 (open square). 

A second CPMG measurement on sample~1 would be required to quantitatively compare the attenuation induced by both CPMG and RMS sequences. 
The CPMG decay rate at $G$=0 can be tentatively inferred from that of sample~2, since conditions were comparable in both sets of experiments. 
The dashed line in Fig. \ref{pentes} is obtained assuming these two rates to be equal (diffusion-independent decay rate due, e.g., to sequence imperfections). 
The dotted line is obtained by appropriate scaling assuming, instead, that decay rates are solely due to diffusion-induced attenuation (in residual inhomogeneities). 
Actual CPMG decay rates are expected to lie between those two boundaries. 
In the following subsection, they are compared to RMS decay rates measured using sample~1.

\subsubsection{Discussion}
\label{diffusionattenuation}
\label{Discussion}
The slope of each line in Fig.~\ref{pentes} directly measures the diffusion coefficient for the corresponding helium mixture (cf. Eq.~\ref{Tanner}). 
As expected, the linear fit to RMS data and the lines demarcating CPMG decay rate limits in sample~1 have very similar slopes (they only differ by 1.5\% for the dotted line and 8.5\% for the dashed line). 
Direct comparison of RMS and CPMG decay rates at fixed $^3$He molar fraction is in progress. 
The RMS decay rates in sample~1 yield $D$=$2.7\times 10^{-3}$~cm$^2$/s for $X$=1.9\%. 
The diffusion coefficient measured with the CPMG sequence in sample~2 is smaller ($D$=$1.3\times 10^{-3}$~cm$^2$/s for $X$=4.5\%). 
Both values agree with published data \cite{opfer68}.

At zero gradient, the echo decay rate is higher for the RMS sequence than for the CPMG sequence. 
The physical origin of this additional damping has not yet been identified. Evaporation of $^3$He atoms due to rf heating (of order 1~mK/s during RMS experiments) has a negligible influence on signal decay. Additional signal losses incurred during RMS sequences, that are observed to be
$B_{\textrm{dip}}$-independent, may be caused by minor rf amplitude and timing imperfections that will soon be addressed. 

\subsection{Repeated magic sandwiches in the presence of dipolar interactions}
\label{Dipolar}
\begin{figure}[ht]
\begin{center}
\includegraphics[width=11cm]{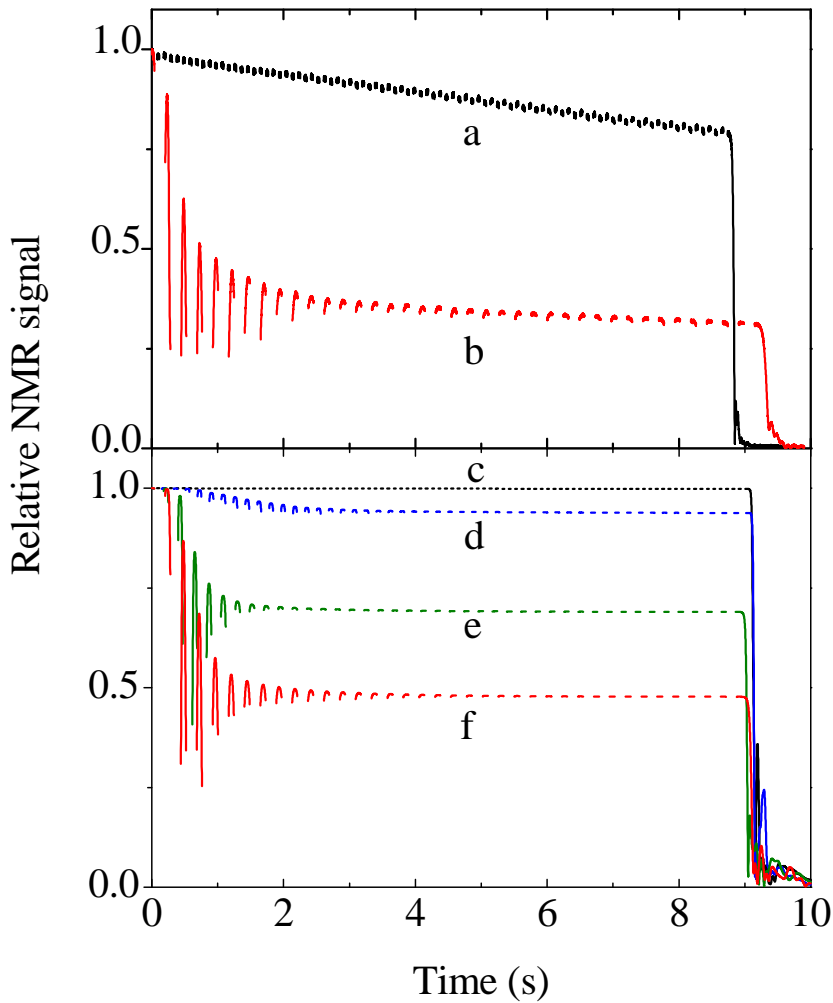}
\caption{\label{RMS} Experimental and simulated RMS echo trains at zero gradient. 
Upper panel: Experiments (traces a and b) are performed with $T_{\textrm{RMS}}$=97~ms and 241~ms, respectively, for identical conditions: $B_{\textrm{dip}}$=0.9~$\mu$T, $X$=4.5\%, and $T$=1.12~K. 
Lower panel : Numerical simulations (traces c to f) are performed using the experimental value of $B_{\textrm{dip}}$, but a larger diffusion coefficient: $D$=$6.4\times 10^{-3}$~cm$^2$/s instead of $D$=$1.3\times 10^{-3}$~cm$^2$/s (see text). The MS repetition time increases from trace c to trace f, with $T_{\textrm{RMS}}$= 97, 180, 217, and 241~ms for these computations.
}
\end{center}
\end{figure}

The upper panel of Fig. \ref{RMS} displays two echo trains recorded for RMS experiments performed at zero gradient on highly polarized samples. 
It illustrates the spectacular change in time behavior of the signal amplitudes when the MS repetition time is modified. 
For the short $T_{\textrm{RMS}}$, the echo amplitude slowly and steadily decreases. 
For the long $T_{\textrm{RMS}}$, the echo amplitude rapidly decreases during the first seconds, then decays as slowly as for the short $T_{\textrm{RMS}}$. 
Dynamic stabilization of transverse magnetization by rf-driven time reversal yields a tremendous increase in precession lifetime: 40~s is measured for the slow signal decays, which is three orders of magnitude larger than the 70~ms obtained for a FID at the same $B_{\textrm{dip}}$ (Fig. \ref{Exponential}, trace b). 
This stabilization results from fast switching between free evolution and rf-driven evolution, i.e., from the frequent sign inversion of the evolutionary effect of dipolar fields. 
The initial rapid decrease observed at long $T_{\textrm{RMS}}$ is reminiscent of the strong DDF-induced attenuation observed at high magnetizations in the single-shot experiments with MS duration $2\tau$ (Fig. \ref{MS}). 
Finally, when the RMS pulse sequence is stopped (about 9~s after the initial 90$^{\circ}$ pulse), signal amplitudes abruptly decrease as they do for simple FIDs following 90$^{\circ}$ tipping angle pulses (cf. Fig.~\ref{Exponential}). 

The lower panel of Fig. \ref{RMS} displays RMS echo trains obtained by numerical simulation. We use a general-purpose model to compute NMR dynamics based on an exact calculation of the time evolution of coupled magnetic moments on a cubic lattice \cite{nacher02,theseMarion}. 
All processes that contribute to the evolution of $\vec M$ are taken into account, including diffusion, dipolar interactions, rf field, and coupling to the detection coil (radiation damping). 
However, this coarse-grained description of a continuous fluid has obvious limitations when spatial variations at the scale of the lattice constant are involved. 
Computations are performed in the rf rotating frame using a standard secular approximation. 
The magnetic field induced at each site by the remainder of the sample is efficiently computed by toggling between real and Fourier spaces \cite{enss99}. 
The time evolution is computed by integrating the nonlinear Bloch equation using a standard Runge-Kutta technique. 
Free evolution usually involves slow local changes, and computations are quite fast on a standard PC computer. 
When intense rf fields are applied, the induced precession is fast and the computing load is significantly increased (e.g., by two orders of magnitude for a magic sandwich). Results presented in Fig. \ref{RMS} have been obtained using $16 \times 16 \times 16$ sites, with computation times of order 1000 s per second of physical evolution time.

For these exploratory simulations, the diffusion coefficient $D$ has been deliberately increased fivefold with respect to the experimental value to efficiently damp spatial variations of $\vec M$ at scales smaller than the lattice constant, as required to obtain meaningful results.
To obtain traces c and f, input parameters $B_{\textrm{dip}}$ and $T_{\textrm{RMS}}$ have been set to their experimental values for traces a and b, respectively. $T_{\textrm{RMS}}$ has been set to intermediate values for traces d and e (see caption).  
The simulated echo trains exhibit the same qualitative features as the experimental data: an initial rapid decrease of echo amplitudes (for all but the shortest $T_{\textrm{RMS}}$), a plateau, and an abrupt signal decrease following the last magic sandwich. 
 
The plateau regime corresponds to the dynamic stabilization of transverse magnetization obtained in the experiments, but the observed slow decay of echo amplitudes is not reproduced. 
A flat plateau is obtained for all simulations. 
The computed magnetization maps show that $\vec M$ is almost uniform, which explains why no noticeable attenuation occurs despite the use of a large diffusion coefficient. The experimental slow decays must thus result from other effects (cf. Sect. \ref{diffusionattenuation}). 
The amplitude of the plateau increases when $T_{\textrm{RMS}}$ is decreased (lower panel), as experimentally observed (upper panel and ref. \cite{hayden07}). It also increases when $D$ is increased (data not shown). This indicates that a higher transverse magnetization can be dynamically stabilized when the development of unstable magnetization patterns is more efficiently prevented. 
A reliable computation of the evolution of $\vec M$ during the first seconds will require using a larger number of sites to more accurately describe the development of inhomogeneities at small length scales.

\section{Conclusion}

NMR dynamics in dipolar-coupled liquid systems can be efficiently investigated with MS-based pulse sequences. 
Using single-shot or multi-echo techniques, the rapid collapse of average transverse magnetization due to the onset of DDF-induced instabilities can be avoided. Dynamic stabilization of precession is achieved by periodic rf-driven time reversal. Experimental and numerical tools have been developed for quantitative investigations of the time evolution of magnetization. These tools will be used to elucidate the potential contribution of dipolar couplings to signal loss, through the development of inhomogeneous magnetization patterns.

\end{document}